# Observation of spin-phonon coupling in GdCl$_3$ filled carbon nanotubes

S. Ncube[1], R. Erasmus[2] and S. Bhattacharyya[1,2,3*]

1  Nano-Scale Transport Physics Laboratory, School of Physics, University of the Witwatersrand, Johannesburg, South Africa
2  DST-NRF Centre of Excellence in Strong materials, University of the Witwatersrand, Johannesburg, South Africa
3  National University of Science and Technology "MISiS", 119049, Leninsky pr. 4, Moscow, Russia

The enhancement of spin-phonon coupling in gadolinium filled double walled carbon nanotubes (GdCl$_3$@DWNTs) is demonstrated through temperature dependent magnetrometry and polarized Raman Spectroscopy (PRS). Temperature dependent susceptibility measurements show that the GdCl$_3$@DWNTs undergo a superparamagnetic phase transion at temperatures below 100 K with a blocking temperature at approximately 47 K. From the temperature dependent PRS, it is observed that a distinct phonon hardening occurs for the G-band modes at temperatures close to the onset of the superparamagentic phase. This is in contract to pristine DWNTs where phonon modes saturates to a constant frequency at low temperatures. The abrupt increase in phonon frequency is indicative of a pronounced spin-phonon coupling. The enhanced coupling in the GdCl$_3$@DWNTs is believed to occur due to resonance between nanotube phonon mode and spin flip frequency as temperature decreases.

Keywords: Raman spectroscopy, carbon nanotubes, spin-phonon coupling, superparamagnetism

**Introduction**

Spin-phonon coupling in carbon nanotubes (CNTs) has attracted much interest due to its potential use in quantum technologies including quantum mechanical resonators, novel qubits and spintronic devices [1] [2] [3]. Although most studies have explored coupling the spin of itinerate electrons to the vibrational modes, there are several studies that have investigated composite systems where spin has been introduced through attachment of molecular magnets or through ferromagnetic leads [4][5]. One area that has however remained largely unexplored for this area of research is the filling the nanotubes with spin active materials. Furthermore, filling nanotubes rather than molecular attachment is advantageous because the CNT also acts as a protecting layer and can greatly increase life span as well as stability of such devices. It has already been demonstrated that nanotubes can be filled with a wide variety of materials and thus it is possible to selectively create a composite system with enhanced spin properties. For example, lanthanide based nanocomposites are known to possess a strong spin-orbit coupling due to the unpaired electrons in the 4f orbitals, this can give rise to exotic physics [6] [7] [8]. Thus understanding magnetization dynamics in low dimensional lanthanide systems is important from a fundamental point of view but may also be useful for technological advancements [8]. Therefore numerous subsequent works have reported on the magnetic, electrical and structural properties of the trivalent lanthanide gadolinium (Gd$^{3+}$) nanocomposite [9] [10] [11]. Although Raman spectroscopy, has to date proven valuable for characterization of lanthanide based composites, there are a few detailed studies of the exchange interaction between magnetic lanthanide ions and their hosts as well as their effect on the phonon modes [9] [12]. Previous studies in magnetic systems have confirmed that spin-phonon interaction can be verified through temperature dependence of the phonon frequencies and linewidths [13][14] [15][16][17][18], and although Raman spectroscopy studies have before been conducted on Gd filled CNTs, to date the effects of the correlation between spin centres and vibrational modes of the nanotubes has not been firmly established.

In this work a nanocomposite consisting of a double walled carbon nanotube (DWNT) which hosts a hexagonal lanthanide salt, gadolinium chloride (GdCl$_3$@DWNT), is studied. The GdCl$_3$@DWNT system has before shown to be superparamagnetic [19], which is in itself an interesting magnetic phase where isolated nanoscopic particle behave as macrospin centers with a temperature dependent spin relaxation rate. Furthermore, as the physics of the phonon modes of DWNTs is well established [17], such a composite system presents an ideal scenario to investigate the possible effects of spin-phonon coupling.

Here we report how the encapsulation of Gd$^{3+}$ ions into the inner core of a CNT changes the physical properties of the resulting nanocomposite. As the filling of CNTs with non-magnetic materials does not affect the properties of the G-band, this phonon mode is of importance for determining the effect of spin-phonon coupling without the need for discriminating between pressure or inter-tube coupling induced effects [14]. Polarized Raman

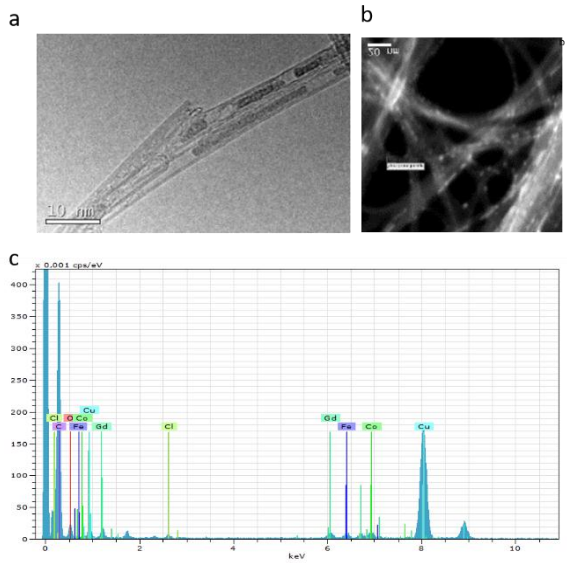

**Fig 1:** (a) HRTEM of GdCl$_3$@DWNTs, the nanoscale GdCl$_3$ can clearly be identified as discontinuous as rod-like structures (b) HAADF image of the filled DWNTs (c) EDX spectra of the nanocomposite confirming the pressence of GdCl$_3$ through the Gd and Cl peaks. Also observed are traces of the catalystic material Co.

Spectroscopy (PRS) is used to differentiate the line shapes of the G-band in pristine and filled DWNTs. Temperature dependant Raman spectroscopy was performed to study the intrinsic correlation between the lattice vibration and the shift induced by the spin-phonon interaction. Magnetic property studies were conducted to further corroborate these findings. This work will improve the understanding of the contribution of spin relaxation due to spin-phonon coupling in one-dimensional systems.

**Results and Discusssion**

DWNTs were prepared using a catalytic chemical vapor depostion technique, the capillary filling method described in [18] was used to incorporate the GdCl$_3$ into the tubes. Both pristine as well as filled tubes were characterized through high resolution transmisson microscopy (HRTEM), energy dispersive spectroscopy (EDX) and Inductively Coupled Plasma Optical Emission Spectrometry (ICP-OES). HRTEM imaging of the GdCl$_3$@DWNTs (Fig 1a) clearly show the successful filling of the tubes where the GdCl$_3$ can be identified as discontinuous rod like structures within the core of the innermost tube. The length of the GdCl$_3$ structures were determined to be in the range of approximately 5 - 30 nm. Through the ICP-OES it was established that the GdCl$_3$@DWNTs had an elemental composition of Co: 1.21 %, Mo: 0.48%, Gd: 6. 90% weight percentage. Thus, any spin related effects are expected to result from the GdCl$_3$ instead of catalystic impurities because of the stronger magnetic moment as well as the higher elemental concentration.

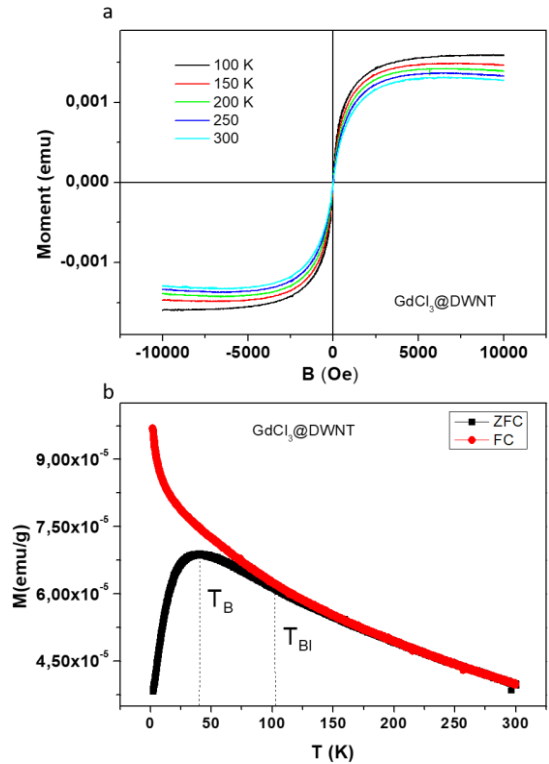

**Fig 2:** (a) Moment as a function of applied field for the GdCl$_3$@DWNTs indicate a paramagnetic signature at all temperatures. (b) Temperature dependent magnetic suscepitbility plot of GdCl$_3$@DWNTs with (FC) and without magnetic field applied (ZFC). The deviation between the two plots ($T_{BI}$ = 100 K) as well as the inflection point ($T_B$ = 47 K) in the ZFC sweep are strong indication of superparmagnetism.

In order to evaluate the effect of the incorporation of a magnetic species into the DWNTs, magnetization measurements were performed in a superconducting quantum unit interference device (SQUID) magnetometer at temperatures between 100 - 300 K. As shown in figure 2(a), the response of the magentic moment with sweeping the applied field between -1000 and 1000 Oe clearly indicates a paramagnetic behaviour with no remenant field at any temeprature between 100 K and 300 K. This is in stark contrast to pristine DWNTs which are predominantly diamagnetic as expected [19].

Magnetic susceptibility was measured in both field cooled (FC) and zero-field cooled (ZFC) conditions from room temperature down to 2 K. As seen in figure 2 b, upon decreasing temperatures the two suceptability curves start to diverge at approximatrly 100 K with the ZC sweep showing an increase in susceptability and ZFC sweep showing a broad inflection point at roughly 47 K and then decreasing rapidly at lower tempertures. These features are characterisitc of a superparamagnetic phase where the deviation between the FC and ZFC sweeps is called the bifurcation temperatue and the inflection point of the ZFC sweep is called the blocking temperture caused by the spin-flip rate of the magnetic diomains reaching a comparable value to that of the measurement time. The broadness of the inflection point is a result fo a distribution in size of the nano-scale magnetic domains. The spin-relaxation thus exhbits a broad frequency range which is dependent on the temepature of the system and although superparamagnetism is deifned by non-interacting domains, when embedded into a host material, it is interting to establish how interactions between the host lattice and magnetic domain can affect the spin relaxation. This is particualrly intersting for the present system as carbon nanotubes have been widely explored as spin-phonon coupled systems. This is largely becuse of the well established physics of the phonon modes in carbon nanotubes and thus the present system gives an ideal platform to investigate the spin-phonon coupling.

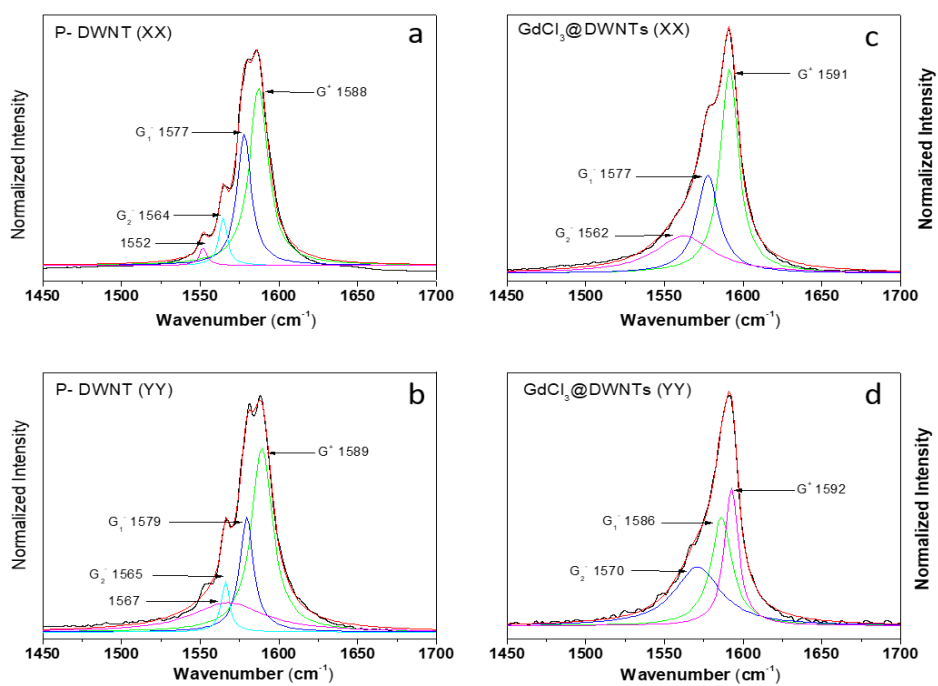

**Fig 3:** Polarized Raman spectrum of the pristine and GdCl$_3$@DWNTs. (a) Deconvolution of the G-band of the pristine DWNTs in the XX and (b) YY configuration shows the G-band can be described by at least four sub modes. For the GdCl$_3$@DWNTs however only three sub bands can be identifed for both (c) XX and (d) YY configuration. Following convention we denote the higher frequency peak as G$^+$ and lower frequency modes as G$_1^-$ and G$_2^-$ as shown above.

In the following a detailed PRS study is conducted to establish spin-phonon coupling in this composite system. In order to correctly correlate the polarization angle used in the PRS study with the tube axis dielectrophoresis (DEP) was used to align GdCl$_3$@DWNT ropes across gold electrodes. This allows for a defined orientation of the nanotubes during the PRS study [20]. All Raman spectra were obtained in backscattering configuration using an excitation wavelength of λ = 514.5 nm (2.41 eV).
PRS was recorded at different orientations denoted XX and YY for horizontal and vertical polarization respectively. As the frequency of spin relaxation is strongly temperature dependent, Raman spectroscopy was conducted at temperatures in a wide temperature range (77 - 300 K). As shown in the inset of figure 3(a&b), the three main sets of peaks can be observed in the Raman spectra of both pristine DWNT [21] and GdCl$_3$@DWNTs. These correspond to the radial breathing modes (RBM), D- and G- bands. As the G -band and RBM are most sensitive to polarization studies and the effects of the filling, we focus on these bands for the present study.

As indicated in figure 3(a-d), there is a clear difference between the G-band profile of pristine DWNTs and GdCl$_3$@DWNTs. Firstly, the G band of the GdCl$_3$@DWNTs is shifted up to 1591 cm$^{-1}$ in comparison to 1588 cm$^{-1}$ for the pristine samples. Secondly, there is a distinct difference in the G peak line shape of the two samples. Upon deconvolution it is seen that there are four sub peaks associated with the pristine material and only three sub peaks associated with the GdCl$_3$@DWNTs. These main sub peaks are denoted as the G$^+$ (higher frequency) and G$^-$ (lower frequency) peaks and correspond to the modes of the inner and outer tubes. The G-band modes are also intimately related to the electronic character of the tubes and are frequently used to discriminate between metallic and semiconducting types, and thus are ideal phonon modes for investigating possible spin effects. For a comparative study we focus on the three main peaks as shown in fig 3(a-d), i.e. G$^+$, G$_1^-$ and G$_2^-$.

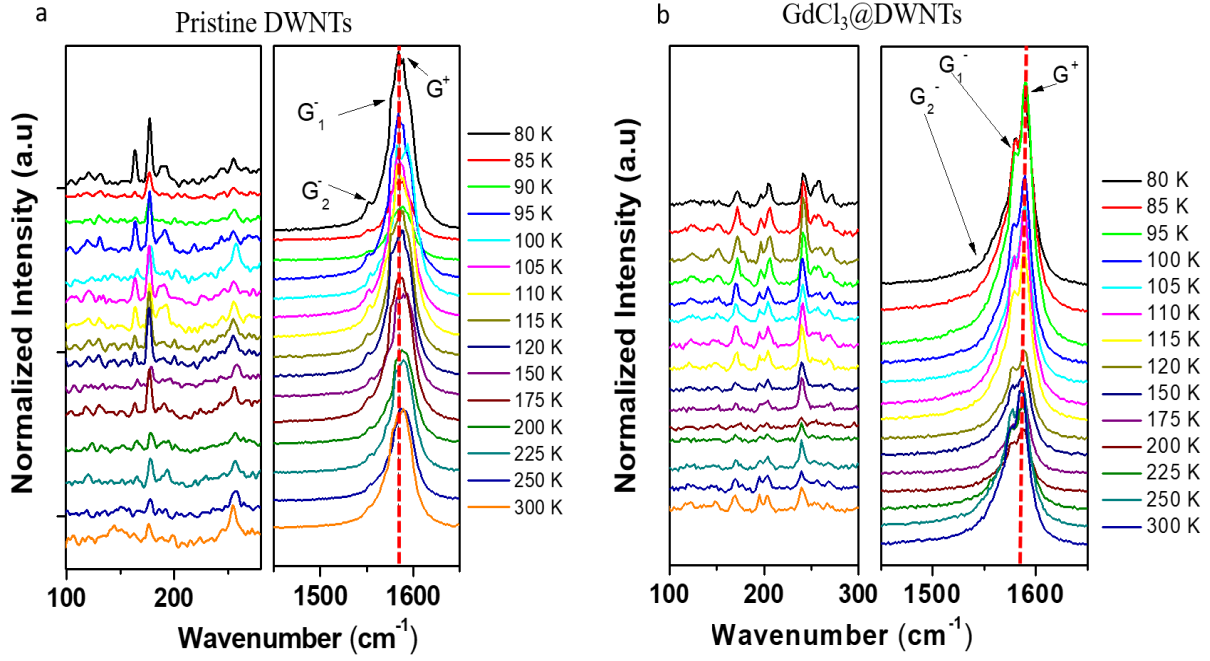

**Fig 4**: Temperature dependent polarized Raman spectra in the XX orientation for both (a) pristine and (b) filled DWNTs. Plotted are both the radial breathing modes as well as the G-band. When comparing the RBMs of the respective samples it is clear that the GdCl$_3$@DWNTs have appreciably mode small frequency peaks, believed to result from the GdCl$_3$ vibrational modes. The G band of both samples clearly show a temperature dependence where the G-band is shifted to higher frequencies as temperature is lowered. The temperature depend shifting is observed in all G-band modes as indicated in the figure.

Indicated in Fig 4(a & b) are the temperature dependent evolution of both the low wavenumber RBM and G- band modes. As expected for DWNTs, both sets of tubes show the characteristic RBMs comprised of multiple peaks of small intensity between 100 and 400 cm$^{-1}$ [16] [21], however, as seen in figure 4(b), the GdCl$_3$@DWNTs show a range of additional low wavenumber peaks which are a signature of the Gd filling [6] and can have up to four bands between 93-230 cm$^{-1}$, attributed to hexagonal structure of GdCl$_3$ [22] [23].

Fig 4(a-b) clearly shows that there is a distinct temperature dependence of the G band for both the filled and the pristine DWNTs. In order to analyze the temperature dependence of Raman peaks in depth, the peak positions of the respective components of the G band are monitored with decreasing temperature. Temperature dependent behavior of the phonon frequency, ω if affected by at least three major terms and is given by [24];

$$\omega(T) = \omega(0) + \Delta\omega_{qh}(T) + \Delta\omega_{sp-lat}(T) + \Delta\omega_{el-ph}(T) \quad (1)$$

where $\omega(0)$ is the frequency at zero K, $\Delta\omega_{qh}(T)$ gives the intrinsic anharmonic contribution between nearest neighbor atoms, $\Delta\omega_{sp-lat}(T)$ represents the change in phonon frequency due to spin-phonon coupling and $\Delta\omega_{el-ph}(T)$ arises takes into account electron-phonon coupling [25].

However, in systems with negligible spin–phonon coupling, the temperature dependence of the phonon modes is given by the Balkanski model [26] which takes into account the decay of the optical phonons purely based on the anharmonic coupling term [15]:

$$\omega(T) = \omega(0) + C\left[1 + \frac{2}{e^{(\hbar\omega_0/2K_BT)}-1}\right] \quad (2)$$

where, $\omega(0) = $ 1597 cm$^{-1}$ and 1591 cm$^{-1}$ are the respective values for filled and pristine samples. C is related to the real part of the phonon's self-energy and equals C = -6 and -11 for filled and pristine respectively. The Balkanski model can also be used to describe the temperature dependence of the linewidth full width at half maximum (FWHM) [26];

$$\Gamma(T) = \Gamma(0) + A\left[1 + \frac{2}{e^{(\hbar\omega_0/2K_BT)}-1}\right] \quad (3)$$

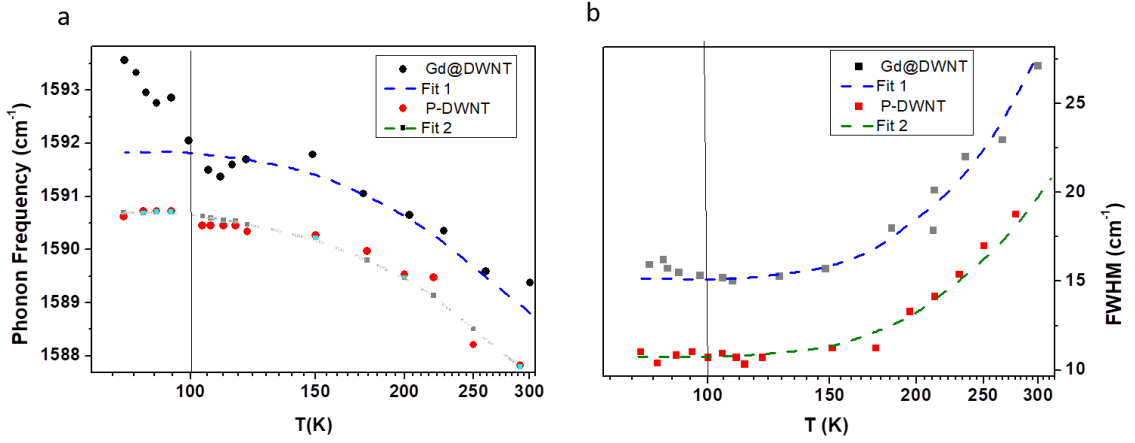

**Fig 5:** (a) A comparison between the G$^+$ phonon frequencies oof the pristine and GdCl$_3$@DWNTs. The phonon frequencies of the pristine samples follow the Balkinski model (dashed line) which is a result a temperatreu decay purely based on anharmonic contributions. The filled samples however show an abrupt deviation fromthis trend at temperatures below 100 K, an indication of spin-phonon coupling (b) The FWHM of the GdCl$_3$@DWNTs however do not how any deviation for the theoretical model, indicating that although spin-phonon occurs in the composite sample, the phonon lifetime is not appreciably affected by the interaction

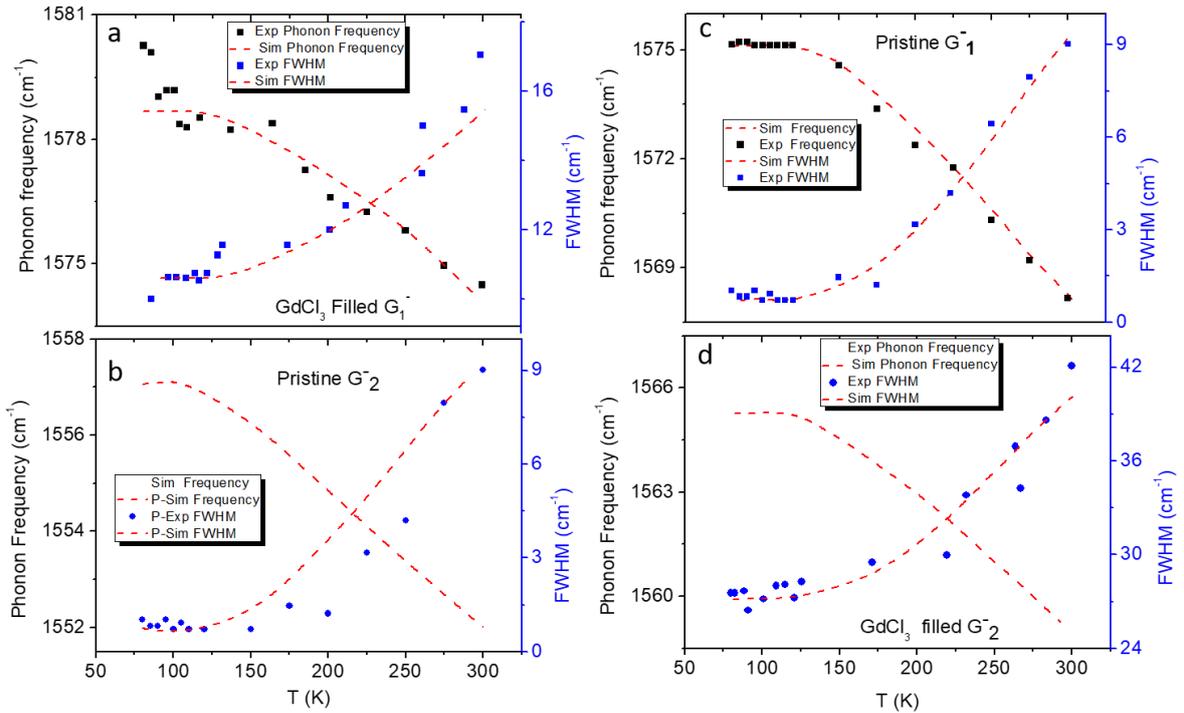

**Fig 6:** The lower frequency sub bands (G$_1^-$ and G$_2^-$) also show a distince temperature dependence, where (a & b) pristine samples completely follow the Blakansi model in strong contract to the filled samples (c&d) where a clear deviation is observed. Intersitngly the magnitude of the frequency shift for the lower frequency modes are smaller than for the higher frequency G$^+$ peak (figure 5(b)).

where $\Gamma(0)$ =15.44 and 10.80 for filled and pristine CNTs respectively, is the value of the FWHM at absolute zero. A is a constant = 1.17 and 0.82 respectively, and the other terms are as defined in equation 2. As the Balkinski model does not account for the spin-phonon coupling, a deviation from the model has been used to identify pronounced spin-phonon coupling along a given temperature range in materials with magnetic phases [13] [15] [27].

As shown in figures 5 and 6, the temperature dependence of the frequency and FWHM of the G$^+$, G$_1^-$ and G$_2^-$ bands have been plotted as a function of the temperature. There is a clear difference between pristine and GdCl$_3$@DWNT material, particualrly in peak frequency temperature dpendence. In the pristine DWNTS case the peak frequency of all G bands neatly follow the theoretical Balkanski model, an indication of neglegable or no spin-phonon interaction. However, in the case of the GdCl$_3$@DWNTs, the peak freqeuncy of the G bands initially follow the Balkinski model but upon lowering temperature show a pronounced deviation from the theoretical trend and increase substantially, indicative of enhanced spin-phon coupling. Furthermore this deviation from the Balkinski model occurs a apporximately 100 K, which was determined to be the bifurcation temperature marking the onset of the superparamagentic state in the magnetic suscetavility data (figure 2(b)). As shown in figures 5(a), 6(b&d), the deviation in peak frequency from the theory is evident in all three G-band modes (i.e. G$^+$, G$_1^-$ and G$_2^-$ ), but is however more pronounced in the G$^+$-band. As this higher frequency mode is associated with the inner tube, a likely explaination of this is the enhanced proximity of tube wall to spin domain as compared to the outer tube modes. This increase in phonon frequency is observed to be as large as 2 cm$^{-1}$, i.e. almost double that reported for materials showing a magnetic ordering phase transitions [13] [15] [27]. Furthermore, as this pronounced increase has not been observed in other filled CNT systems [15] it is unlikely that filling induced pressure or inter wall coupling leads to this observation. Hence the presence of 6.90 % Gd is the most probable cause of the deviations observed in Figure 4 and 5 due to its high magnetic moment and anisotropy.

The effect of the spin-phonon coupling on the FWHM is however not as pronounced as with the peak freqeuncy, as shown in figures 5(b) and 6(a&b), where the FWHM of the modes in both filled and GdCl$_3$@DWNTs do not show much deviation from the Balkinski model. However, as the FWHM is related to the phonon lifetime, the enhanced spin-phonon coupling does not act as a phonon relaxation mechanism and rather leads to phonon hardening (hence increase in the peak frequency). What is remarkable about the observation here is that it is correlated to the superparamagentic phase. As the tenmperature of the system is lowered, the Neel relaxation time (time between spin-flip events) increases, and thus the spin-flipping of the nano-magnetic GdCl$_3$ has a characteristic frequency which increases with decreasing temperature. At low temperatures when the spin relaxation is appreciably low and comparable to the frequency of the phonon modes, resonance occurs and the spin-phonon coupling between localized spin dominas and the phonons of the nanotubes are coupled. This is particualrly interesting when considering the spin relaxation mechanisms known for the superparamagnetic breakdown. At higher temperatures the spin relaxation is considered to be dominated by phonon interference and characterized by an arhenius type depenedence, however at lower temperatures in the superparamagetic phase, spin relaxation is beleived to occur through quantum mechnaical tunnelling of the magnetic moment due to weak particle interaction [28] [29]. Here we have demonstrated that even through there is pronounced spin-phonon coupling, the superparamagnetic phase is not compromised as would be expected. Thus, the GdCl$_3$@DWNTs system offers a unique situation where information about isolated nano-scale localized spin centres can be obtained directly from the vibrational modes of the host nanotubes without disturbing the spin states. As carbon nanotubes have been well studied as potential matter for quantum technology, this study further highlights their suitability for novel electronic device application when filled with magnetic materials.

**Conclusion**

Through the encapsulation of GdCl$_3$ by DWNTs, we have demonstrated an enhanced spin-phonon coupling. It can be seen that the GdCl$_3$@DWNTs system exhibits a superparamagnetic phase which occurs at 100 K. This temperature happens to also coincide with the abrupt increase in G-band phonon frequency and the FWHM, a hallmark feature of spin-phonon coupling. These results provide an insight for the understanding of the spin-phonon interaction in one-dimensional systems which can be important for novel carbon based spintronic application.

*Acknowledgements:* We acknowledges URC Wits, NRF, and CSIR-NLC for funding the project and C. Coleman for useful discussions. We thank E. Flahaut for the sample and some characterization and AM Strydom for the SQUID measurements. SB acknowledges useful discussions with A. Irjhak and financial support from Open Int. grant competition of NUST-MISiS, Moscow.